\documentclass[11pt,onecolumn]{IEEEtran}

\usepackage{amsmath}
\usepackage{amssymb}
\usepackage{amsmath}
\usepackage{amsthm}
\usepackage{amssymb}
\usepackage{amscd}
\usepackage{latexsym}
\usepackage{multirow}
\usepackage{graphics}
\usepackage{color}
\usepackage{cite}
\usepackage{mathtools}
\usepackage{amsfonts}
\usepackage[font=footnotesize]{caption}
\usepackage{setspace}
\usepackage{flushend}
\usepackage{caption}
\usepackage{subcaption}

\newcommand{\ma}[1]{\mathbf{#1}}

\newcommand{\tr}[1]{\mathrm{Tr}\left(#1\right)}
\newcommand{\sgn}[1]{\mathrm{sgn}\left(#1\right)}
\newcommand{\oma}[1]{\ma #1_{\text{\normalfont opt}}}

\newcommand{\mxmz}{\mathrm{maximize}}

\newcommand{\maximize}[2]{\underset{\substack{#1}}{\mxmz}~#2}

\newcommand{\maxm}[2]{\underset{\substack{#1}}{\max}~#2}

\newcommand{\reals}{\mathbb R}

\date{}
\newcommand{\eqdef}{\stackrel{\cdot}{=}}

\newtheoremstyle{mynewtheorem}
{5pt}
{5pt}
{\it}
{}
{\bf}
{.}
{.5em}
{\thmname{#1}\thmnumber{#2}}%

\theoremstyle{mynewtheorem}
\newtheorem{myprop}{Proposition }
\newtheorem{mylemma}{Lemma }

\newtheorem{myassum}{Assumption }

\usepackage{graphicx}
\usepackage{subcaption}
\DeclareCaptionLabelFormat{r-parens}{{#2}} 
\begin{document}
	
	\title{The Exact Solution to Rank-1 L1-norm TUCKER2 Decomposition}

	\author{
		\thanks{P. P. Markopoulos and D. G. Chachlakis are with the Department of Electrical and Microelectronic Engineering, Rochester Institute of Technology, Rochester, NY 14623 USA (e-mail: \texttt{panos@rit.edu, dimitris@mail.rit.edu}).
			
			E. E. Papalexakis is with the Department of Computer Science and Engineering, University of California Riverside, Riverside, CA 92521 USA (e-mail: \texttt{epapalex@cs.ucr.edu}).	
		}
		Panos P. Markopoulos,$^*$\thanks{\hrule$^*$Corresponding author.}
		Dimitris G. Chachlakis, 
		and Evangelos E. Papalexakis
	}
	
\markboth{THIS IS A PREPRINT; AN EDITED/FINALIZED VERSION OF THIS  MANUSCRIPT HAS BEEN SUBMITTED TO IEEE SIGNAL PROCESSING LETTERS}{Markopoulos et al.: {The Exact Solution to Rank-1 L1-norm TUCKER2 Decomposition}}

	\maketitle

\begin{center}
	\vspace{-2cm}
\today	
\end{center}

	\begin{abstract}
		We study rank-1 {L1-norm-based TUCKER2} (L1-TUCKER2) decomposition of 3-way tensors, treated as a collection of $N$ $D \times M$ matrices that are to be jointly decomposed. Our contributions are as follows.
		 i) We prove that the problem is equivalent to combinatorial optimization over $N$ antipodal-binary variables. ii) We derive the first two algorithms in the literature for its exact solution. The first algorithm has cost exponential in $N$; the second one has cost polynomial in $N$ (under a mild assumption). Our algorithms are accompanied by formal complexity analysis. iii) We conduct numerical studies to compare the performance of exact L1-TUCKER2 (proposed) with standard HOSVD, HOOI, GLRAM, PCA, L1-PCA, and TPCA-L1. Our studies show that L1-TUCKER2 outperforms (in tensor approximation) all the above counterparts when the processed data are outlier corrupted.  
	\end{abstract}
	

	\section{Introduction and Problem Statement}
	\label{problem}
    {I}{ntroduced} by L. R. Tucker \cite{TUCKER} in the mid-1960s, TUCKER decomposition is a fundamental method $n$-way tensor analysis, with applications in a wide range of fields, including machine learning, computer vision \cite{VASILESCU,Lathauwer}, wireless communications \cite{GOMEZ}, biomedical signal processing \cite{biomed}, and social-network data analysis \cite{SUN0,Kolda0} to name a few.
	Considering that the $n$-way tensor under processing is formed by the concatenation (say, across the $n$-th mode, with no loss of generality) of a number of coherent (same class, or distribution) $(n-1)$-way coherent tensor measurements, then TUCKER decomposition simplifies to TUCKER2 decomposition. 
	TUCKER2 strives to \emph{jointly} decompose the collected $(n-1)$-way tensors and unveil the low-rank multi-linear structure of their class, or distribution.
    Higher-Order SVD (HOSVD) and Higher-Order Orthogonal Iteration (HOOI) algorithms \cite{Lathauwer2} are well-known solvers for TUCKER2 (and TUCKER) decompositions. 
	A detailed presentation of TUCKER, TUCKER2, and the respective solvers is offered in \cite{Kolda,SIDIROPOULOS,SUN}. 
	Note that both types of solvers can generally only guarantee a locally optimal solution.
	
	For $n=2$, TUCKER/TUCKER2 take the familiar form of Principal-Component Analysis (PCA). 
	Thus, similar to PCA, TUCKER/TUCKER2 are sensitive against outliers within the processed tensor \cite{PANG,CAO,FU}.
	On the other hand, L1-Principal-component Analysis (L1-PCA) \cite{PPM0,PPM1,PPM2}, substituting the L2-norm in PCA by the outlier-resistant L1-norm, has illustrated remarkable outlier-resistance. 
	Extending this formulation to tensor processing, one can similarly endow robustness to TUCKER and TUCKER2 decompositions by substituting the L2-norm in their formulations by the L1-norm. 
	Indeed, an approximate algorithm for L1-norm-based TUCKER2 (L1-TUCKER2) was proposed in \cite{PANG}.
	However, L1-TUCKER2 remains to date unsolved.
	In this work, we offer for the first time the exact solution to L1-TUCKER2 for the special case of rank-1 approximation, and provide two optimal algorithms. 
	A formal problem statement follows.
	
	Consider a collection of $N$ real-valued matrices of equal size, $\ma X_{1}, \ma X_2, \ldots, \ma X_{N} \in \reals^{D \times M}$.
	For any rank $d \leq \min\{ D, M \}$, a \emph{TUCKER2} decomposition strives to jointly analyze $\{ \mathbf X_{i}\}_{i=1}^N$, by maximizing $\sum_{i=1}^N \| \ma U^\top \ma X_i \ma V \|_{F}^2$ over $
	\ma U \in \reals^{D \times d}$ and $ \ma V \in \reals^{M \times d}$, such that $~\ma U^\top \ma U = \ma V^\top \ma V = \ma I_{d}$; then, $\mathbf X_{i}$ is low-rank approximated as $\mathbf U\mathbf U^\top \mathbf X_{i} \mathbf V \mathbf V^\top$. 
	The squared Frobenius norm $\|\cdot\|_F^2$ returns the summation of the squared entries of its matrix argument. 
	Among other methods in the tensor-processing literature, TUCKER2 coincides with  Multilinear PCA \cite{LU} (for zero-centered matrices) and the {Generalized Low-Rank Approximation of Matrices (GLRAM)} \cite{YE}.
	Clearly, for $N=1$, TUCKER2 simplifies to the rank-$d$ approximation of  matrix $\ma X_1 \in \mathbb R^{D \times M}$, solved by means of the familiar singular-value decomposition (SVD) \cite{GOLUB}; i.e., the optimal arguments $\ma U$ and $\ma V$ are built by  the $d$ left-hand and right-hand singular vectors of $\ma X_1$, respectively.
	
	To counteract against the impact of any outliers in $\{ \mathbf X_{i}\}_{i=1}^N$, in this work, we consider the L1-norm-based TUCKER2 reformulation
	\begin{align}
	\text{L1-TUCKER2:~~~}
	\maximize{
		\begin{smallmatrix}
		\ma U \in \reals^{D \times d};~\ma U^\top \ma U = \ma I_{d} \\
		\ma V \in \reals^{M \times d};~\ma V^\top \ma V = \ma I_{d} \\
		\end{smallmatrix}
	}
	{ \sum_{i=1}^N \| \ma U^\top \ma X_i \ma V \|_1},
	\label{l1glram}
	\end{align}
	where the L1-norm $\|\cdot\|_1$ returns the summation of the absolute values of its matrix argument. 
	The problem in \eqref{l1glram} was studied in \cite{PANG} under the title L1-Tensor Principal-Component Analysis (TPCA-L1).\footnote{ In this work, we refer to the problem as L1-TUCKER2, so as to highlight its connection with the TUCKER2 formulation (instead of the  general TUCKER formulation).}
	Authors in \cite{PANG} presented an approximate algorithm for its solution which they employed for image reconstruction. To date,  \eqref{l1glram} has not been solved exactly in the literature, even for the special case of rank-1 approximation (i.e., $d=1$). 
	In this work, we deliver, for the first time,  the exact solution to L1-TUCKER2 for $d=1$, by means of two novel algorithms. In addition, we provide numerical studies that demonstrate the outlier-resistance of exact L1-TUCKER2, and its superiority (in joint-matrix decomposition and reconstruction) over L2-norm-based (standard) TUCKER2, GLRAM, TPCA-L1, PCA, and L1-PCA.

	\section{Exact Solution}

	\subsection{Reformulation into combinatorial optimization}
	For rank $d=1$, L1-TUCKER2 in \eqref{l1glram} takes the form 
	\begin{align}
	\maximize{
		\begin{smallmatrix}
		\ma u \in \mathbb R^{D \times 1};~\ma v \in \mathbb R^{M \times 1};~\| \ma u \|_2=\| \ma v \|_2=1
		\end{smallmatrix}
	}
	{ \sum_{i=1}^N | \ma u^\top \ma X_i \ma v|}
	\label{l1glram1}
	\end{align}
	First, we focus on  the absolute value in \eqref{l1glram1} and notice  that, for any  $\ma a \in \reals^{N}$,   $\sum_{i=1}^N|a_i| = \sum_{i=1}^N \sgn{a_{i}} a_i = \sgn{\ma a}^\top \ma a = \max_{\mathbf b \in \{\pm 1\}^N} \ma b^\top \ma a$, where $\sgn{\cdot}$ returns the  $\{\pm 1\}$-sign of its (vector) argument.  In view of the above,  Lemma 1 follows.
	\begin{mylemma}
		For  any given $\mathbf u \in \reals^{D}$ and  $\mathbf v \in \reals^M$, it holds that 
		\begin{align}
		{ \sum_{i=1}^N | \ma u^\top \ma X_i \ma v|}  &= \maxm{\mathbf b \in \{\pm 1 \}^N }{~ \ma u^\top \left(  \sum_{i=1}^N b_{i} \ma X_i \right) \ma v }. 
		\label{eqlem1}
		\end{align}
		The maximum  in  \eqref{eqlem1} is attained for  $\ma b = [\sgn{\ma u^\top \ma X_{1} \ma v}, $ $ \sgn{\ma u^\top \ma X_{2} \ma v},$ $\ldots, \sgn{\ma u^\top \ma X_{N} \ma v}]^\top$. \hfill $\square$
	\end{mylemma}
	In addition, the following well-known Lemma 2 derives by the matrix-approximation optimality of SVD \cite{GOLUB}. 
	\begin{mylemma}
		For  any given $\mathbf b \in \{ \pm 1 \}^N$,  it holds that 
		\begin{align}
		\maxm{
			\begin{smallmatrix}
			\ma u \in \reals^{D \times 1};~ \| \ma u \|_2=1 \\
			\ma v \in \reals^{M \times 1};~ \| \ma v \|_2=1 \\
			\end{smallmatrix}
		}\hspace{-0.4cm}
		{   \ma u^\top \left( \sum_{i=1}^N b_{i} \ma X_i \right) \ma v }   
		=  \sigma_{\max}\left( \sum_{i=1}^N b_{i} \ma X_i \right)
		\label{eqlem2}
		\end{align}
		where $\sigma_{\max}(\cdot)$ returns the highest singular value of its matrix argument. 
		The maximum  in  \eqref{eqlem2} is attained if $\mathbf u$ and $\mathbf v$ are the left-hand and right-hand dominant singular vectors of $\sum_{i=1}^N b_{i} \ma X_i$, respectively. \hfill $\square$
	\end{mylemma}
	
	To compact our notation, we  concatenate $\{ \ma X_{i} \}_{i=1}^N$ into  $\mathbf X \eqdef [\mathbf X_{1}, \mathbf X_{2}, \ldots, \mathbf X_{N}] \in \mathbb R^{D \times MN}$. Then,  for any $\mathbf b \in \{ \pm 1 \}^{N}$, it holds $\sum_{i=1}^{N} b_{i} \mathbf X_{i} = \ma X (\mathbf b \otimes \mathbf I_{M})$, where $\otimes$ denotes the Kronecker matrix product \cite{VANLOAN}.
	Then, in view of Lemma 1 and Lemma 2, we can rewrite the L1-TUCKER2 in \eqref{l1glram1} as 
	\begin{align}
	& \maxm{
		\begin{smallmatrix}
		\ma u \in \reals^{D \times 1};~ \| \ma u \|_2=1 \\
		\ma v \in \reals^{M \times 1};~ \| \ma v \|_2=1 \\
		\end{smallmatrix}
	}
	\sum_{i=1}^N   |\ma u^\top \ma X_i \ma v | \label{way1} \\
	= &  
	\maxm{
		\begin{smallmatrix}
		\ma b \in \{ \pm  1\}^{N} \\
		\ma u \in \reals^{D \times 1};~ \| \ma u \|_2=1 \\
		\ma v \in \reals^{M \times 1};~ \| \ma v \|_2=1 
		\end{smallmatrix}
	}
	\ma u^\top \left(\ma X (\ma b \otimes \ma I_{M}) \right) \ma v  
	\label{way3} \\
	= &  
	\maxm{\ma b \in \{ \pm  1\}^{N}} 
	\sigma_{\max} \left( \ma X (\ma b \otimes \ma I_{M}) \right). 
	\label{way4}
	\end{align}
	It is clear that \eqref{way4} is a combinatorial problem over the size-$2^N$ feasibility set $\{ \pm  1\}^{N}$.  The following Proposition 1 derives straightforwardly from Lemma 1, Lemma 2, and \eqref{way1}-\eqref{way4} and concludes our transformation of \eqref{l1glram1} into a combinatorial problem. 
	\begin{myprop}
		Let $\oma b$ be a solution to the combinatorial 
		\begin{align}
		\maximize{\mathbf b \in \{ \pm \}^N }{ \sigma_{\max}(\mathbf X(\mathbf b \otimes \mathbf I_{M}))}
		\label{binary}
		\end{align}
		and denote by $\oma u \in \reals^{D}$ and $\oma v \in \reals^{M}$ the left- and right-hand singular vectors of $\mathbf X(\oma b \otimes \mathbf I_{M}) \in \reals^{D \times M}$, respectively. Then, $(\oma u, \oma v)$ is an optimal solution  to \eqref{l1glram1}. Also,  $\oma b = [\sgn{\oma u^\top \ma X_{1} \oma v}, $ $ \ldots,  \sgn{\oma u^\top \ma X_{N} \oma v}]^\top$  and $\sum_{i=1}^N | \oma u^\top \ma X_i \oma v| = \oma u^\top (\mathbf X(\oma b \otimes \mathbf I_{M}))  \oma v $ $= \sigma_{\max}\left( \mathbf X(\oma b \otimes \mathbf I_{M})\right)$. In the special case that $\oma u^\top \ma X_{i} \oma v=0$, for some $i \in \{1, 2, \ldots, N\}$, $[\oma b]_{i}$ can be set to $+1$, having no effect to the metric of \eqref{binary}.
		\hfill $\square$ 
	\end{myprop}

	Given $\oma b$, $(\oma u, \oma v)$ are obtained by SVD of $\mathbf X(\oma b \otimes \mathbf I_{M})$. Thus, by Proposition 1, the solution to L1-TUCKER2 for low-rank $d=1$ is obtained by the solution of the combinatorial problem \eqref{binary} and a $D$-by-$M$ SVD.

	\subsection{Connection to L1-PCA and hardness}
	
	In the sequel, we show that for $M=1$ and $d=1$, L1-TUCKER2 in \eqref{l1glram1}  simplifies to L1-PCA \cite{PPM0,PPM1,PPM2}. Specifically, for $M=1$, matrix $\mathbf X_{i}$ is a $D \times 1$ vector, satisfying   $\mathbf X_{i} = \mathbf x_{i} \eqdef \text{vec}(\mathbf X_{i})$, and \eqref{l1glram1} can be rewritten as 
	\begin{align}
	\maxm{
		\begin{smallmatrix}
		\ma u \in \reals^D;~v \in \reals;~\| \ma u\|_2=|v|=1 
		\end{smallmatrix}
	}{\sum_{i=1}^N | \ma u^\top \mathbf x_{i} v|}.
	\label{M1}
	\end{align} 
	It is clear that for every $\ma u$, an optimal value for $v$ is trivially $v=1$ (or, equivalently, $v=-1$); thus, for $\mathbf X = [\mathbf x_{1}, \mathbf x_{2}, \ldots, \mathbf x_{N}] \in \mathbb R^{D \times N}$,  \eqref{M1} becomes 
	\begin{align}
	\maxm{
		\begin{smallmatrix}
		\ma u \in \reals^D; ~\| \ma u\|_2=1 
		\end{smallmatrix}
	}{\sum_{i=1}^N | \ma u^\top \mathbf x_{i}|}
	=
	\maxm{
		\begin{smallmatrix}
		\ma u \in \reals^D; ~\| \ma u\|_2=1 
		\end{smallmatrix}
	} \|  \mathbf X^\top \ma u \|_1,
	\label{M1_2}
	\end{align} 
	which is the exact formulation of the well-studied L1-PCA problem \cite{PPM0,PPM1,PPM2}. 
	We notice also that for $M=1$  the combinatorial optimization \eqref{binary} in Proposition 1 becomes 
	\begin{align}
	\maxm{\mathbf b \in \{ \pm 1\}^N}{ \sigma_{\max} (\mathbf X (\mathbf b \otimes 1) ) } =
	\maxm{\mathbf b \in \{ \pm 1\}^N}{ \| \mathbf X \mathbf b \|_2 }, 
	\label{binary2}
	\end{align}
	since the maximum singular-value of a vector coincides with its Euclidean norm, which is in accordance to the L1-PCA analysis in \cite{PPM1,PPM2}. 
	Based of the equivalence of L1-PCA to  \eqref{binary2}, \cite{PPM1} has proven that L1-PCA of $\mathbf X$ is formally \emph{NP}-hard in $N$, for jointly asymptotic $N$ and $\text{rank}(\mathbf X)$. Thus, by its equivalence to L1-PCA for $d=1$ and $M=1$, L1-TUCKER2 is also \emph{NP}-hard in $N$, for jointly asymptotic $N$ and $\text{rank}(\mathbf X)$.


	\subsection{Exact Algorithm 1: Exhaustive search}
	
	Proposition 1 shows how the solution to \eqref{l1glram1} can be obtained through the solution to the combinatorial problem in \eqref{binary}. Our first exact algorithm solves \eqref{binary} straightforwardly by an exhaustive search over its feasibility set. In fact, noticing that $\sigma_{max}(\cdot)$ is invariant to negations of its matrix argument, we obtain a solution $\oma b$ to \eqref{binary} by an exhaustive search in the size-$2^{N-1}$ set $\mathcal B_{\text{ex}} = \{ \ma b \in \{ \pm 1\}^N: ~b_1 =1 \}$. For every value that $\mathbf b$ takes in $\mathcal B_{\text{ex}}$, we conduct SVD to $\ma X(\mathbf b \otimes \ma I_{M})$ to calculate $\sigma_{\max}(\ma X(\mathbf b \otimes \ma I_{M}))$, with cost $\mathcal O(\min\{D,M\} DM )$ \cite{GOLUB}. 
	Since it entails $2^{N-1}$ SVD calculations, the cost of this exhaustive-search algorithm is $\mathcal O(2^{N-1} \min\{D,M\} DM)$; thus, it is exponential to the number of jointly processed matrices, $N$, and at most quadratic to the matrix sizes, $D$ and $M$.  

	\begin{figure}[t]
		\centering
		\includegraphics[draft=false,scale=.75]{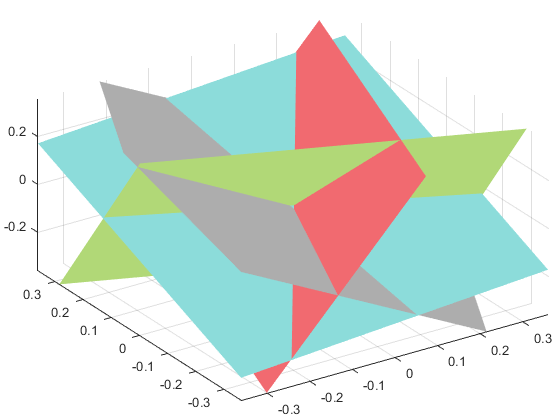}
		\caption{For $\rho=3$ and $N=4$, we draw $\mathbf W \in \mathbb R^{\rho \times N}$, such that  $\mathbf W \mathbf W^\top = \mathbf I_{3}$ and Assumption 1 holds true. Then, we plot the nullspaces of all $4$ columns of $\mathbf W$ (colored planes). We observe that the planes partition $\mathbb R^{3}$ into $K=2(\binom{3}{0} + \binom{3}{1} + \binom{3}{2}) = 2(1+3+3)=14$ coherent cells  (i.e., $7$ visible cells above the cyan hyperplane and $7$ cells below.) \vspace{-0.2cm}}
		\label{fig:hpl}
	\end{figure}

	\subsection{Exact Algorithm 2: Search with cost polynomial in $N$}
	
	In  the sequel, we focus on the case where $N$ is low-bounded by the constant $DM$ and present an algorithm that solves \eqref{l1glram1} with polynomial cost in $N$. $DM < N$ emerges as a case of interest in signal processing applications when $\{ \mathbf X_{i} \}_{i=1}^N$ are measurements of a $D \times M$ fixed-size sensing system (e.g., $D \times M$ images). 
	By Proposition 1, for the optimal solutions $\oma b$ and $(\oma u, \oma v)$ of \eqref{binary} and \eqref{l1glram1}, respectively, it holds 
	\begin{align}
	\oma b & = [\sgn{\oma v^\top \ma X_{1}^\top \oma u}, \ldots, \sgn{\oma v^\top \ma X_{N}^\top \oma u} ]^\top,
	\label{uv2b}
	\end{align} 
	with
	$
	\sgn{\oma u^\top \ma X_{i} \oma v} = +1, \text{~if~} \oma u^\top \ma X_{i} \oma v=0.
	$
	In addition, for every $i \in \{1, 2, \ldots, N\}$, we find that
	\begin{align}
	\oma v^\top \ma X_{i}^\top \oma u &= \tr{\ma X_{i}^\top \oma u \oma v^\top} = \mathbf x_{i}^\top (\oma v \otimes \oma u). 
	\end{align}
	Therefore, defining $\ma Y = [\ma x_{1}, \ma x_{2}, \ldots, \ma x_{N}] \in \mathbb R^{DM \times N}$,  \eqref{uv2b} can be rewritten as 
	\begin{align}
	\oma b = \sgn{ \ma Y^\top (\oma v \otimes \oma u)}.
	\label{uv2b2}
	\end{align}
	Consider now that $\ma Y$ is of some rank $\rho \leq \min \{ DM, N \}$ and admits  SVD $\ma Y \overset{svd}{=} \mathbf Q \ma S \ma W$, where $\mathbf Q^\top \mathbf Q = \mathbf W \mathbf W^\top = \mathbf I_{\rho}$ and  $\ma S$ is the $\rho \times \rho$  diagonal matrix that  carries  the $\rho$ non-zero singular-values of $\ma Y$.
	Defining  $\oma p \eqdef \ma S^\top \ma Q^\top(\oma v \otimes \oma u)$, \eqref{uv2b2} can be rewritten as
	\begin{align}
	\oma b & = \sgn{ {\ma W}^\top \oma p}.
	\label{uv2b2}
	\end{align} 
	In view of \eqref{uv2b2} and since $\sgn{\cdot}$ is invariant to positive scalings of its vector argument, 
	an optimal solution to \eqref{binary}, $\oma b$, can be found in the binary set
	\begin{align}
	\mathcal B =\{ \mathbf b \in \{ \pm 1\}^N:~\mathbf b=  \sgn{ {\ma W}^\top \ma c}, ~\ma c \in \mathbb R^{\rho} \}.
	\label{bset}
	\end{align}
	Certainly, by definition, \eqref{bset} is a subset of $\{\pm 1\}^{N}$ and, thus, has finite size  upper bounded by $2^N$.
	This, in turn, implies that there exist instances of $\mathbf c \in  \mathbb R^{\rho}$ that yield the same value in $ \sgn{ {\ma W}^\top \ma c}$. Below, we delve into this observation to  build a tight superset of $\mathcal B$ that has polynomial size in $N$, under the following mild ``general position" assumption \cite{YALE}. 
	\begin{myassum}
		For  every $\mathcal I \subset \{ 1, 2, \ldots, N \}$ with $|\mathcal I|=\rho-1$, it holds that $\text{rank}( [\mathbf W]_{ :, \mathcal I})=\rho-1$; i.e., any collection of $\rho-1$ columns  of $\mathbf W$ are linearly independent.	
	\end{myassum}

	For   any $i \in \{1, 2, \ldots, N\}$, define $\ma w_{i} \eqdef [\mathbf W]_{:,i}$ and denote by $\mathcal N_i$ the nullspace of $\ma w_i$. Then, for every $ \ma c \in \mathcal N_i$, the (non-negative) angle between $\mathbf c$ and $\mathbf w_{i}$,  $\phi(\mathbf c, \mathbf w_{i})$, is equal to $\frac{\pi}{2}$ and, accordingly,  $\ma w_i^\top \mathbf c =  \|\mathbf c \|_2 \|\mathbf w_{i} \|_2 \cos{\left(\phi(\mathbf c, \mathbf w_{i}) \right)} =0$. Clearly, the \emph{hyperplane}  $\mathcal N_{i}$ partitions $\mathbb R^{\rho}$ in two non-overlapping \emph{halfspaces}, $\mathcal H_{i}^+$ and $\mathcal H_i^{-}$ \cite{ORLIKTERAO}, such that $\sgn{\mathbf c^\top \mathbf w_{i}} = +1$ for every $\mathbf c \in \mathcal H_{i}^+$ and $\sgn{\mathbf c^\top \mathbf w_{i}} = -1$ for every $\mathbf c \in \mathcal H_{i}^-$. 
	In accordance with Proposition 1, we consider that $\mathcal H_{i}^+$ is a  closed set that includes its boundary $\mathcal N_i$, whereas $\mathcal H_{i}^-$ is open and does not overlap with $\mathcal N_i$. 
	In view of these definitions, we proceed with the following illustrative example. Consider some $\rho > 2$ and two column indices  $m <i  \in \{1, 2, \ldots, N\}$. Then, hyperplanes $\mathcal N_m$ and $\mathcal N_i$ divide  $\mathbb R^{\rho}$ in the halfspace pairs  $\{ \mathcal H_{m}^{+}, \mathcal H_{m}^{-} \}$ and $\{ \mathcal H_{i}^{+}, \mathcal H_{i}^{-} \}$, respectively.  By Assumption 1,\footnote{If $\mathbf w_{m}$ and $\mathbf w_{i}$ are linearly independent, then $\mathcal N_m$ and $\mathcal N_i$ intersect but do not coincide.}  each one of the two halfspaces defined by $\mathcal N_m$ will intersect with both  halfspaces defined by $\mathcal N_i$, forming the four halfspace-intersection ``cells"  $\mathcal C_1 = \mathcal H_{m}^+ \cap \mathcal H_{i}^{+}$, $\mathcal C_2 = \mathcal H_{m}^+ \cap \mathcal H_{i}^{-}$, $\mathcal C_3 = \mathcal H_{m}^- \cap \mathcal H_{i}^{-}$, $\mathcal C_4 = \mathcal H_{m}^- \cap \mathcal H_{i}^{+}$. 
	It is now clear that, for any $k \in \{1, 2, 3, 4\}$,  $[\sgn{ [\mathbf W]^\top \mathbf c}]_{m,i}$ is the same for every $\ma c \in \mathcal C_k$. For example,   for every $\mathbf c \in \mathcal C_2$,  it holds that $[\sgn{ [\mathbf W]^\top \mathbf c}]_m = +1$ and $[\sgn{ [\mathbf W]^\top \mathbf c}]_i = -1$.

	Next, we go one step further and consider the arrangement of all $N$ hyperplanes $\{ \mathcal N_{i} \}_{i=1}^N$. Similar to our discussion above, these hyperplanes partition $\mathbb R^{\rho}$ in  $K$ cells  $\{ \mathcal C_k\}_{k=1}^K$, where $K$ depends on $\rho$ and $N$. Formally, for every $k$, the $k$-th halfspace-intersection set is defined as
	\begin{align}
	\mathcal C_{k} \eqdef \bigcap_{i \in \mathcal I_{k}^{+}} \mathcal H_{i}^{+} \bigcap_{m \in \mathcal I_{k}^{-1}} \mathcal H_{m}^{-},
	\label{coh} 
	\end{align}
	for  complementary index sets $\mathcal I_{k}^+$ and $\mathcal I_{k}^-$ that satisfy $\mathcal I_{k}^+ \cap \mathcal I_{k}^- = \emptyset$ and $\mathcal I_{k}^+ \cup \mathcal I_{k}^- = \{1, 2, \ldots, N\}$  \cite{WINDER,BROWN}. 
	By the definition in \eqref{coh},  and in accordance with our example above, every $\mathbf c  \in \mathcal C_{k}$ lies in the same intersection of halfpsaces and, thus, yields the exact same value in $\sgn{\mathbf W^\top \mathbf c}$. Specifically, for every $\mathbf c  \in \mathcal C_{k}$, it holds that
	\begin{align}
	\left[\sgn{\mathbf W^\top \mathbf c}\right]_{i} = 
	\sgn{\mathbf w_i^\top \mathbf c} =
	\left\{ 
	\begin{matrix}
	+1, & i \in \mathcal I_{k}^{+}  \\
	-1, & i \in \mathcal I_{k}^{-}  
	\end{matrix}
	\right..
	\label{prop}
	\end{align}
	In view of \eqref{prop}, for every $k \in \{1,2, \ldots, K\}$ and any  $\ma c \in \mathcal C_k$, we define the ``signature" of the $k$-th  cell  $\ma b_k \eqdef \sgn{\mathbf W^\top \mathbf c}$.
	Moreover,  we observe that $\mathcal C_k \cap \mathcal C_l = \emptyset$ for every $k \neq l$ and that $\cup_{k=1}^K \mathcal C_k= \mathbb R^{\rho}$. By the above observations and definitions, \eqref{bset} can be rewritten as
	\begin{align}
	\mathcal B & = \bigcup_{k=1}^K \{   \sgn{ {\ma W}^\top \ma c}: ~\ma c \in   \mathcal C_k \} = \{ \ma b_1, \ma b_2, \ldots, \ma b_K\}.
	\label{bset2}
	\end{align}
	Importantly, in \cite{WINDER,KARYSTINOS}, it was shown that the exact number of coherent cells formed by the nullspaces of $N$ points in $\mathbb R^{\rho}$ that are in general position (under Assumption 1) is exactly 
	\begin{align}
	K = 2 \sum_{j=0}^{\rho-1} \binom{N-1}{j} \leq 2^N,
	\label{exact}
	\end{align}
	with equality in \eqref{exact} if and only if $\rho = N$.
	Accordingly, per \eqref{exact}, the cardinality of $\mathcal B$ in \eqref{bset} is  equal to $|\mathcal B|= 2 \sum_{j=0}^{\rho-1} \binom{N-1}{j}$. 
	For clarity,  in Fig. \ref{fig:hpl}, we plot the nullspaces (colored planes) of the columns of arbitrary $\mathbf W \in \mathbb R^{3 \times 4}$ that satisfies both $\mathbf W \mathbf W^\top =\ma I_3$ and Assumption 1. It is interesting that exactly $K=14 < 2^{4} =16$  coherent cells emerge  by the intersection of the formed halfspaces. 
	In the sequel, we rely on \eqref{bset2} to develop a conceptually simple method for computing  a tight superset of the cell signatures in $\mathcal B$.

	
	Under Assumption 1,   for any  $\mathcal I \subseteq \{1,2, \ldots, N \}$ with $|\mathcal I|=\rho-1$, the  hyperplane intersection $\mathcal V_{\mathcal I} \eqdef \cap_{i\in \mathcal I} \mathcal N_i$  is a line ($1$-dimensional subspace) in $\mathbf R^{\rho}$. By its definition, this line is the verge between all cells that are jointly bounded by the $\rho-1$ hyperplanes in $\{ \mathcal N_i \}_{i \in \mathcal I}$. Consider now a vector $\ma c \in \mathbb R^{\rho}$ that crosses over the verge $\mathcal V_{\mathcal I}$ (at any point other than $\ma 0_{\rho}$). By this crossing, the value of  $[\sgn{\mathbf W^\top \ma c}]_{\mathcal I}$ will change so that $\sgn{\mathbf W^\top \ma c}$ adjusts to the signature of the new cell to which $\ma c$ just entered. At the same time, a crossing over $\mathcal V_{\mathcal I}$ cannot be simultaneously  over any of the hyperplanes in $\{\mathcal N_{i}\}_{i \in \mathcal I^c}$, for $\mathcal I^c \eqdef \{ 1,2, \ldots, N\} \setminus \mathcal I$; this is because, under Assumption 1, it is only at $\ma 0_{\rho}$ that more than $\rho-1$ hyperplanes can intersect.
	Therefore, it is clear that  $[\sgn{\mathbf W^\top \ma c}]_{\mathcal I^c}$ will remain invariant during this crossing and, in fact,  equal to  $ [\sgn{\mathbf W^\top \ma v}]_{\mathcal I^c}$, for any $\ma v \in \mathcal V_{\mathcal I}$ with $\ma v^\top \ma c>0$.  
	In view of the above, for any  $\ma v \in \mathcal V_{\mathcal I}\setminus \ma 0_{\rho}$, the set
	\begin{align}
	\mathcal B_{\mathcal I} \eqdef \{ \ma b \in \{\pm 1\}^N: [\ma b]_{\mathcal I^c} =  [\sgn{\mathbf W^\top  \ma v}]_{\mathcal I^c} \}
	\label{BI}
	\end{align}
	contains the signatures of all sets that are bounded by the verge $\mathcal V_{\mathcal I}$.
	Moreover, it has been shown (see, e.g., \cite{KARYSTINOS}) that, for every  cell, there exists at least one such verge that bounds it. Therefore, it derives that the set   
	\begin{align}
	\mathcal B_{\text{pol}} = \underset{
		\begin{smallmatrix}\mathcal I \subset \{1, 2, \ldots, N \};~
		|\mathcal I|=\rho-1
		\end{smallmatrix}
	}{\bigcup} \mathcal B_{\mathcal I}
	\end{align}
	includes all cell signatures and, thus, is a superset of $\mathcal B$. We notice that, for every $\mathcal I$,  $\mathcal B_{\mathcal I}$ has size $2^{\rho-1}$. Since $\mathcal I$ can take $ \binom{N}{\rho-1}$ distinct values, we find that $\mathcal B_{\text{pol}}$ is upper bounded by $2^{\rho-1}\binom{N}{\rho-1}$. Thus, both $|\mathcal B_{\text{pol}}|$ and $|\mathcal B|$ are polynomial, in the order of $\mathcal O(N^{\rho-1})$. 
	
	Practically, for every $\mathcal I$,  $\ma v$ can be calculated by Gram-Schmidt orthogonalization of $[\ma W]_{:,\mathcal I}$ with cost $\mathcal O(\rho^{3})$. Keeping the dominant terms, the construction of $\mathcal B_{\text{pol}}$ costs $\mathcal O(N^{\rho-1})$ and can be parallelized in $\binom{N}{\rho-1}$ processes. Then, testing every entry of $\mathcal B_{\text{pol}}$ for optimality in \eqref{binary} costs an additional $\mathcal O(N)$. Thus, the overall cost of our second algorithm, taking also into account the $\mathcal O(N)$ (for constant $DM$) SVD cost for the formation of $\ma W$, is $\mathcal O(N^{\rho})$. The presented algorithm is summarized in  Fig. \ref{fig:algo2}.

	\begin{figure}[t!]
		{\small
			{\hrule height 0.2mm} 
			\vspace{0.3mm}
			{\hrule height 0.2mm}
			\vspace{1mm}
			{\bf Algorithm 2: Polynomial in $N$}
			\vspace{0.5mm}
			{\hrule height 0.2mm}
			\vspace{2mm}
			\textbf{Input:} $\{ \mathbf X_i\}_{i=1}^N$  \\
			\begin{tabular}{r l }
				0: & $\mathbf Y \leftarrow [\text{vec}(\mathbf X_{1}),\text{vec}(\mathbf X_{2}), \ldots, \text{vec}(\mathbf X_{N})]$ \\
				1: & $(\mathbf Q, \mathbf S_{d \times d}, \mathbf W) \leftarrow \mathrm{svd} (\mathbf Y)$, $m_t \leftarrow 0$ \\
				2: &For every $\mathcal I \subseteq \{1, 2, \ldots, N\}$, $|\mathcal I|=d-1$ \\
				3: &~~~~~ Build $\mathcal B_{\mathcal I}$ in \eqref{BI} 
				\\
				4: &~~~~~ For every $\ma b \in \mathcal B_{\mathcal I}$  \\
				5: &~~~~~ ~~~~~~ $(\mathbf U, \mathbf \Sigma, \mathbf V) \leftarrow \mathrm{svd} (\mathbf X(\mathbf b \otimes \mathbf I_{M}))$\\
				6: &~~~~~ ~~~~~~ $m \leftarrow \max\{ \text{diag}( \mathbf \Sigma) \}$ \\
				7: &~~~~~ ~~~~~~ if $m>m_t$, \\
				8: &~~~~~ ~~~~~~ ~~~~~~ $m_t \leftarrow m$, $\ma b_{t}\leftarrow \ma b$, $\ma u \leftarrow [\ma U]_{:,1}$, $\ma v \leftarrow [\ma V]_{:,1}$  
			\end{tabular} \\
			\textbf{Output:} $\oma b \leftarrow \ma b_t$, $\oma u \leftarrow \ma u$, and $\oma v \leftarrow \ma v$
			\vspace{1.5mm}
			{\hrule height 0.2mm} 
			\vspace{0.3mm}
			{\hrule height 0.2mm}
			\vspace{0.4cm}
		}
		\caption{Algorithm for the exact solution of rank-1 L1-TUCKER2 in \eqref{l1glram1}, with cost $\mathcal O(N^{\rho+1})$. }
		\vspace{-0.6cm}
		\label{fig:algo2}
	\end{figure}

	\section{Numerical Studies}
	
	Consider $\{\mathbf X_{i}\}_{i=1}^{14}$ such that 
	$
	\mathbf X_{i}={\mathbf A_i} +\ma N_i \in \mathbb R^{20 \times 20}
	$
	where $\ma A_{i}=b_{i} \ma u \ma v^\top $ and $\| \ma u\|_2 = \| \ma v \|_2=1$,  $b_i\sim \mathcal{N}(0, 49)$, and each entry of $\mathbf N_i$ is additive white Gaussian noise (AWGN) from  $\mathcal{N}(0,1)$.
	We consider that $\ma A_{i}$ is the rank-1 useful data in $\ma X_{i}$ that we want to reconstruct, by joint analysis (TUCKER2-type) of $\{\ma X_{i}\}_{i=1}^{14}$.
	By irregular corruption, $30$ entries in $2$ out of the $14$ matrices (i.e., $60$ entries out of the total $5600$ entries in $\{\mathbf X_{i}\}_{i=1}^{14}$) have been further corrupted additively by noise from 
	$\mathcal{N}(0,\sigma_c^2)$. 
	To reconstruct $\{\mathbf A_{i}\}_{i=1}^{14}$ from $\{\mathbf X_{i}\}_{i=1}^{14}$, we follow one of the two approaches below.

	\begin{figure}[t]
		\centering
		\includegraphics[draft=false,scale=.85]{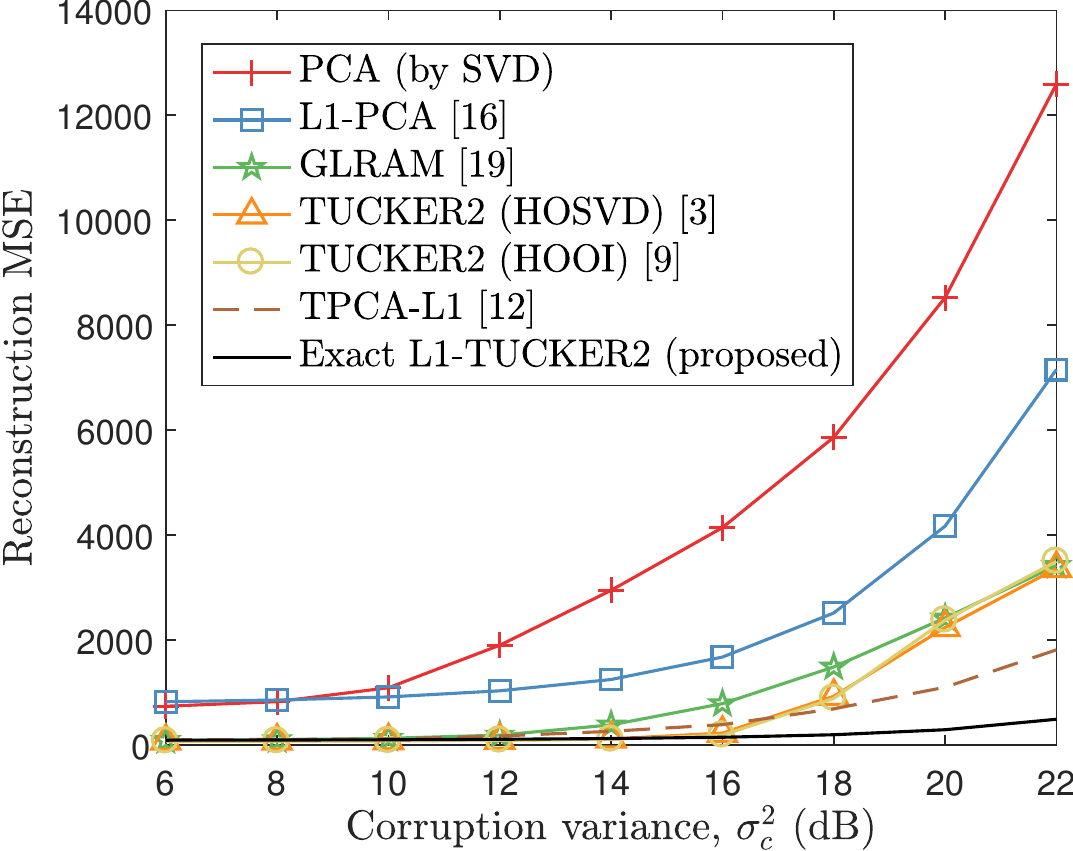}
		\caption{Reconstruction MSE versus corruption variance $\sigma_c^2$ (dB). 
		}
		\label{fig:numerical_study}
	\end{figure}

	In the first approach, we vectorize the matrix samples and perform standard  matrix analysis. 
	That is, we obtain the first ($d=1$) principal component (PC) of $[\text{vec}(\ma X_1), \text{vec}(\ma X_2),\ldots , \text{vec}(\ma X_N)]$,  $\mathbf q$. 
	Then, for every $i$, we approximate $\ma A_i$ by $\hat{\ma A}_i=\text{mat}(\mathbf q \ma q^\top \ma a_i)$, where $\text{mat}(\cdot)$ reshapes its vector argument into a $20 \times 20$ matrix, in  accordance with $\text{vec}(\cdot)$. 
	In the second approach, we process the samples in their natural form, as matrices, analyzing them by TUCKER2.
	If (${\ma u}, {\ma v}$) is the TUCKER2 solution pair, then we approximate $\mathbf A_{i}$ by  $\hat{\ma A}_i=  {\ma u} {\ma u}^\top \ma X_i  {\ma v} {\ma v}^\top$.
	For the first approach, we obtain $\mathbf q$ by PCA (i.e., SVD) and L1-PCA \cite{PPM1}. 
	For the second  approach, we conduct TUCKER2 by HOSVD \cite{Lathauwer}, HOOI \cite{Kolda}, GLRAM \cite{YE}, TPCA-L1 \cite{PANG}, and the proposed exact L1-TUCKER2.
	Then, for each reconstruction method, we measure the mean of the squared error $\sum_{i=1}^{14} \| \mathbf A_{i} - \hat{\mathbf A}_{i} \|_F^2$ over 1000 independent realizations for corruption variance $\sigma_c^2=6,8,\ldots,22$dB.
	In Fig. \ref{fig:numerical_study}, we plot the reconstruction mean squared error (MSE) for every method, versus $\sigma_c^2$.
	We observe that PCA and L1-PCA exhibit the highest MSE due to the vectorization operation (L1-PCA outperforms PCA clearly, across all values of $\sigma_c^2$). 
	Then, all TUCKER2-type methods perform similarly well when $\sigma_c^2$ is low. 
	As the outlier variance $\sigma_c^2$ increases, the performance of L2-norm-based TUCKER2 (HOSVD, HOOI) and GLRAM deteriorates severely. On the other hand, the L1-norm-based TPCA-L1 exhibits some robustness. The proposed exact L1-TUCKER2 maintains the sturdiest resistance against the corruption, outperforming its counterparts across the board.

	\flushend

\end{document}